\documentclass[useAMS,usenatbib,a4paper,preprint]{mn2e}
\usepackage{graphicx}
\usepackage{amsmath}
\usepackage{txfonts}
\usepackage{mathrsfs}

\usepackage{color}

\def\aap{A\&A}
\def\apjl{ApJ}

\def\apjs{ApJS}

\def\apjs{ApJ Supplements}

\def\apjl{ApJ Letters}

\def\aap{A\&A}

\newcommand{\be}{\begin{equation}}
\newcommand{\ee}{\end{equation}}
\newcommand{\bea}{\begin{eqnarray}}
\newcommand{\eea}{\end{eqnarray}}

\title[Speed of light]{Model-independent confirmation of a
constant speed of light over cosmological distances}
\author[Fulvio Melia]{Fulvio Melia\thanks{John Woodruff Simpson
Fellow. E-mail: fmelia@email.arizona.edu}\\
Department of Physics, The Applied Math Program, and Department of Astronomy,
The University of Arizona, AZ 85721, USA}

\begin{document}

\date{}

\pagerange{\pageref{firstpage}--\pageref{lastpage}} \pubyear{2023}

\maketitle

\label{firstpage}

\begin{abstract}
Recent attempts at measuring the variation of $c$ using an assortment
of standard candles and the redshift-dependent Hubble expansion rate
inferred from the currently available catalog of cosmic chronometers
have tended to show that the speed of light appears to be constant, 
at least up to $z\sim 2$. A notable exception is the use of high-redshift
UV $+$ X-ray quasars, whose Hubble diagram seems to indicate a
$\sim 2.7\sigma$ deviation of c from its value $c_0$ ($\equiv 2.99792458
\times 10^{10}$ cm s$^{-1}$) on Earth. We show in this paper, however,
that this anomaly is due to an error in the derived relation between 
the luminosity distance, $D_L$, and $H(z)$ when $c$ is allowed to vary
with redshift, and an imprecise calibration of the quasar catalog. When 
these deficiences are addressed correctly, one finds that $c/c_0=0.95
\pm 0.14$ in the redshift range $0\lesssim z\lesssim 2$, fully consistent 
with zero variation within the measurement errors.
\end{abstract}

\begin{keywords}
{cosmology: theory -- gravitation -- quasars: general}
\end{keywords}

\section{Introduction}\label{intro}
The first known measurement of the speed of light was carried out in
1676 by the Danish astronomer Olaus Roemer \citep{Romer:1940}, who
timed the eclipses of the Jovian moon Io and estimated that the delay
of about 22 minutes seen on diametrically opposite sides of Earth's
orbit around the Sun was due to light's travel time across that
distance. Planetary distances were not well known back then, but
he nevertheless obtained a value of $\sim 2\times 10^{10}$ cm s$^{-1}$,
not too different from the much more precise measurement\footnote{Actually,
this value is exact when quoted in these units because, by a 1983
international agreement, a `metre' is defined in terms of how far
light travels in $1/299792458$ seconds. See
https://www.nist.gov/si-redefinition/definitions-si-base-units}
we have today, $c_0=2.99792458\times 10^{10}$ cm s$^{-1}$. (Throughout
this paper, we shall use the symbol $c_0$ to denote the speed of light 
on Earth, to distinguish it from the possible redshift dependent value, 
$c(z)$, inferred from high-redshift data.)

Many have wondered whether $c$ is constant over much larger distances,
however, perhaps even in time, most critically over cosmic scales. And 
several attempts have already been made to determine whether $c$ is in fact 
a true constant of nature. But the problem with measuring $c$ across the
Universe is that one cannot avoid using the Friedmann-Lema\^itre-Robertson-Walker 
(FLRW) metric to describe the background cosmology. Of course, all solutions to 
Einstein's equations in general relativity assume ab initio that $c$ is 
constant. After all, the interval $ds$ written in terms of the metric
coefficients would not even exist if $c$ were variable in space and/or
time \citep{Melia:2020}. 

So unfortunately one is faced with an inherent inconsistency when attempting 
to {\it measure} the hypothesized variability of $c$ using a metric whose 
validity requires that $c$ be constant. In this paper, we shall not avoid 
this issue either, so our goal will not be to actually measure a possible 
time-dependence (or redshift-dependence) of $c$ but, rather, to demonstrate 
whether the cosmological data self-consistently imply that $c$ is constant, 
as required by FLRW. 

The consequences of a variable $c$ were considered by Einstein himself 
\citep{Einstein:1907}, and many workers have followed suit, including
most prominently \cite{Albrecht:1999}, \cite{Barrow:1999a}, \cite{Barrow:1999b}
and \cite{Bassett:2000}, who proposed that $c$ has varied across the 
evolutionary history of the Universe. Similar ideas were espoused and
strongly supported by \cite{Moffat:2002}, and further developed by others
since then. 

But most cosmological tests of the constancy of $c$ have tended to be 
carried out in the context of $\Lambda$CDM. Some attempts at avoiding
the presumption of a background cosmology have also been made, e.g.,
by \cite{Liu:2023}, who `measured' the speed of light using cosmic 
chronometers to infer the Hubble constant, $H(z)$, along with Type~Ia 
supernovae and the UV-X-ray correlation of high-redshift quasars. 
Their results based on the Type~Ia SNe showed strong consistency with
a constant value for $c$. For the quasars, however, they demonstrated 
that the speed of light up to redshift $z\sim 2$ satisfies the constraint 
$c/c_0=1.19\pm 0.07$, in terms of its local value on Earth, $c_0$. Though
the difference between this $c$ and $c_0$ is less than $3\sigma$, an
inconsistency of $\sim 2.7\sigma$ nevertheless fails to affirm the
supernova result and creates some tension. In addition, as noted earlier, 
one cannot really be satisfied with $c$ varying like this when the
analysis is carried out using the FLRW metric, which was not constructed 
for a variable $c$ in the first place. As it stands, the outcome based
on the use of quasars as standard candles either implies that their presumed 
UV-X-ray correlation is not sufficiently precise for this type of analysis 
(see \S\S~\ref{background} and \ref{data} for more details), or perhaps 
that the assumed cosmic spacetime itself is inaccurate.

There appear to be mitigating factors with this conclusion, however.  First, 
the relation derived by \cite{Liu:2023} for the luminosity distance, $D_L$,
in terms of $H(z)$, misses the contribution from $dc(z)/dz$ itself,
which introduces an inconsistency when the inferred value of $c(z)$
is variable. Secondly, previous work with the UV-X-ray catalog
of high-z quasars \citep{Risaliti:2015,Risaliti:2019,Lusso:2010,Lusso:2016}
has suggested that the calibration of these sources for use as standard
candles is more accurately carried out with the simultaneous optimization
of all the parameters, including those characterizing the cosmology,
be it $\Lambda$CDM or the more generic `chosmographic' polynomial fit,
rather than with the separate introduction of luminosity distances
inferred from Type~Ia supernovae \citep{Melia:2019c}.

This type of analysis overlaps with our previous application of 
high-redshift quasars to model selection, a head-to-head comparison between 
$\Lambda$CDM and an alternative FLRW cosmology known as the $R_{\rm h}=ct$ 
universe \citep{Melia:2007,MeliaShevchuk:2012,Melia:2019c}. This test uses 
a recently refined method of sampling the redshift-distance relationship 
\citep{Risaliti:2015} based on an inferred correlation between 
the X-ray and UV monochromatic quasar luminosities, following an earlier 
proposal by \cite{Avni:1986}. From this study emerged the conclusion
that the use of `external' calibrators rendered the high-$z$ quasar
Hubble diagram marginally inconsistent with the predictions of 
$\Lambda$CDM, while an internal calibration using the simultaneous
optimization of the UV-X-ray correlation relation and the model's
parameters produced greater self-consistency. Given that the Type Ia SNe
produced results consistent with a constant $c$, while the quasar Hubble
diagram did not, the role played by this problem with the calibration of
the UV-X-ray correlation needs to be better understood. 

Our goal in this paper is thus to revisit the use of high-$z$ quasars and
cosmic chronometers to test the constancy of $c$, but this time using
a more valid expression for $dc(z)/dz$, along with a better calibration
of the UV-X-ray correlation function. And in keeping with the spirit
of avoiding cosmology-dependent factors as much as possible, we shall
also restrict our attention to the generic 'cosmographic' polynomial 
fit to the data \citep{Risaliti:2019}, rather than relying on the
parameterization in $\Lambda$CDM. 

We begin with a brief description of the background for this
test in \S~\ref{background}, and then describe the quasar and
cosmic-chronometer data and our method of analysis in \S~\ref{data}.
We end with a discussion of our results and a brief conclusion
in \S~\ref{discussion}.

\section{The Quasar UV-X-ray Correlation}\label{background}
Adopting the cosmological principle, which assumes isotropy
and homogeneity throughout the cosmic spacetime, we may write
the metric using the Friedmann-Lema\^itre-Robertson-Walker
(FLRW) ansatz,
\begin{equation}
ds^2=c^2\,dt^2-a^2(t)\left[{dr^2\over 1-kr^2}+r^2\,d\Omega^2\right]\;,\label{eq:FLRW}
\end{equation}
where $a(t)$ is the expansion factor, $t$ is the cosmic time, and
$(r,\theta,\phi)$ are the spatial coordinates in the comoving frame.
In addition, $d\Omega^2\equiv d\theta^2+\sin^2\theta\,d\phi^2$, and
$k$ is the spatial curvature constant, which we assume to be zero
in accordance with most of the data available to us today \citep{Planck:2016}.

In this context, the luminosity distance may be written as
\begin{equation}
D_L(z) = c(z)(1+z)\int_0^z{du\over H(u)}\;,\label{eq:DL}
\end{equation}
where $H(z)\equiv \dot{a}/a$ is the redshift-dependent Hubble parameter,
and we have explicitly written the speed of light as $c(z)$ to remind ourselves
that derivatives of $c$ cannot be ignored should it turn out to be variable.
Throughout this paper, we shall use the form
\begin{equation}
H(z) \equiv H_0E(z)\;,\label{eq:Hz}
\end{equation}
where $H_0$ is the Hubble constant.

Then,
\begin{equation}
D_L^\prime\equiv{d D_L\over dz}=c^\prime{D_L\over c}+
c\int_0^z{du\over H(u)}+{c(1+z)\over H(z)}\;,\label{eq:dDL1}
\end{equation}
from which we derive
\begin{equation}
{c^\prime\over c}={D_L^\prime\over D_L}-{1\over(1+z)}-{1\over H(z)
\int_0^z du/H(u)}\;.\label{eq:dc}
\end{equation}
Very importantly, note that $H_0$ completely cancels out in
this expression. This is an especially desirable feature of this analysis
because of the growing disparity between the measurements of $H_0$ at low and high 
redshifts, creating a $\sim 4\sigma$ uncertainty in its value \citep{Riess:2021}.

And to further simplify the analysis, we divide Equations~(\ref{eq:DL})
and (\ref{eq:dDL1}) and the numerators and denominators in Equation~(\ref{eq:dc}) by 
the constant $c_0$, yielding the differential equation for the `normalized' speed 
of light $\kappa\equiv c/c_0$:
\begin{equation}
{\kappa^\prime\over \kappa}={\tilde{D}_L^\prime\over \tilde{D}_L}-{1\over(1+z)}-{1\over H(z)
\int_0^z du/H(u)}\;,\label{eq:dckappa}
\end{equation}
where
\begin{equation}
\tilde{D}_L(z) = \kappa(z)(1+z)\int_0^z{du\over H(u)}\;,\label{eq:DLkappa}
\end{equation}
and $\tilde{D}_L^\prime$ is correspondingly derived from $\tilde{D}_L(z)$.
This allows us to solve for $c(z)/c_0$ without having to specify the actual
value of $c_0$ locally. 

Our test of the variation of $c$ with redshift will therefore rely on two
independent cosmological measurements: (i) the luminosity distance $D_L$
inferred from the high-$z$ quasar Hubble diagram \citep{Risaliti:2019}, from 
which we shall also calculate the derivative $D_L^\prime$, and (ii) the 
redshift-dependent Hubble parameter $H(z)$ estimated using cosmic chronometers, 
as originally proposed by \cite{Jimenez:2002} and \cite{Jimenez:2003}.

The quasar's UV (i.e., disk) emission is correlated with its X-ray (i.e., coronal)
emission according to the following parametrization: 
\begin{equation}
\log_{10} L_X=\gamma\log_{10} L_{UV}+\beta\;,\label{eq:LXLUV}
\end{equation}
where $L_X$ and $L_{UV}$ are the rest-frame monochromatic luminosities
at 2 keV and 2,500 \AA, respectively, and $\gamma$ and $\beta$ are two
parameters that we shall optimize simultaneously with those characterizing
our generic cosmographic cosmology described below. We shall use base 10
logarithms throughout this paper. Given that the data contain the fluxes, rather
than model-dependent luminosities, we use a slightly modified form of 
Equation~(\ref{eq:LXLUV}),
\begin{equation}
\log_{10}F_X=\tilde{\beta}+\gamma\log_{10} F_{UV}+2(\gamma-1)\log_{10} D_L\;,\label{eq:FXFUV}
\end{equation}
where the constant $\tilde{\beta}$ subsumes the slope $\gamma$ and intercept $\beta$, 
so that
\begin{equation}
\tilde{\beta}=\beta+(\gamma-1)\log_{10} 4\pi\;.\label{eq:beta}
\end{equation}

\begin{table*}
\begin{center}
\caption{Optimization of the UV-X-ray correlation in high-$z$ quasars}
\begin{tabular}{cccccc}
&&&&& \\
\hline\hline
$\tilde{\beta}$ & $\gamma$ & $\delta$ & $a_2$ & $a_3$ & $\chi^2_{\rm dof}$ \\
\hline
&&&&& \\
$6.249\pm0.02$ & $0.626\pm0.0006$ & $0.231\pm0.001$ & $2.93\pm0.33$ & $2.65\pm0.80$ & $1.003$ \\
&&&&& \\
\hline\hline
\end{tabular}
\end{center}
\end{table*}

The luminosity distance in the cosmographic empirical fit we employ here is
given by the expression
\begin{equation}
D_L=\ln(10){c\over H_0}\left[\log_{10}(1+z)+a_2\log_{10}^2(1+z)
+a_3\log_{10}^3(1+z)\right]\;,\label{eq:DLcosmo}
\end{equation}
based on a third-order polynomial with two constants, $a_2$ and $a_3$, that 
we shall optimize along with the other free parameters \citep{Risaliti:2019},
by minimizing the likelihood function
\begin{equation}
\ln(LF)=-\sum_{i=1}^{1598}\left\{{\left[\log_{10}\left(F_X\right)_i-\Phi\left(
\left[F_{UV}\right]_i,D_L\left[z_i\right]\right)\right]^2\over{\tilde{\sigma_i}}^2} +
\ln\left({\tilde{\sigma_i}}^2\right)\right\}\;.\label{eq:LF}
\end{equation}
In this expression, the variance ${\tilde{\sigma_i}}^2\equiv\delta^2+\sigma_i^2$ is given in
terms of a global intrinsic dispersion, $\delta$, and the individual measurement errors
$\sigma_i$ in $(F_X)_i$ \citep{Risaliti:2015}, while the errors in $(F_{UV})_i$ are 
insignificant compared to $\sigma_i$ and $\delta$, so we ignore them in this application. 
In Equation~(\ref{eq:LF}), $\Phi$ is defined as
\begin{equation}
\Phi\left(\left[F_{UV}\right]_i,D_L\left[z_i\right]\right)\equiv\tilde{\beta}+
\gamma\log_{10}\left(F_{UV}\right)_i+2(\gamma-1)\log_{10} D_L(z_i)\;,\label{eq:Phi}
\end{equation}
incorporating the measured fluxes $(F_X)_i$ and $(F_{UV})_i$ at redshift $z_i$.

Once the values of $a_2$ and $a_3$ have been optimized from the quasar data, we
can also calculate the derivative of $D_L$ from Equation~(\ref{eq:DLcosmo}), yielding
\begin{equation}
D_L^\prime = {c\over H_0(1+z)}\left[1+2a_2\log_{10}(1+z)+3a_3\log_{10}^2(1+z)\right]\;,\label{eq:DLp}
\end{equation}
and we divide both $D_L$ and $D_L^\prime$ by $c_0$ to obtain
$\tilde{D}_L$ and $\tilde{D}_L^\prime$ to be used in Equation~(\ref{eq:dckappa}).

\section{Data and analysis}\label{data}
With their high luminosities, quasars represent promising cosmological probes out to larger 
redshifts than many other types of source. Their luminosity distance appears to follow a 
reliable, nonlinear correlation between their ultraviolet (UV) and X-ray monochromatic 
fluxes (Eq.~\ref{eq:LXLUV}). But although this correlation has been known for over three
decades \citep{Avni:1986}, only recently has the impractically large dispersion in 
this relation been suppressed by refining the selection criteria and flux measurements 
\citep{Risaliti:2015,Risaliti:2019,Lusso:2016}. These improvements allow
the quasars to be used as distance indicators out to redshifts $\sim 6$.

\begin{table}
    \caption{Hubble parameter $H(z)$ from cosmic chronometers}
\begin{tabular}{lcl}
    \hline
    $z$ & $H(z) \ (\mathrm{km}\,\mathrm{s}^{-1}\,\mathrm{Mpc}^{-1}) $ & References \\
    \hline
    0.09 & 69  $\pm$  12   & \cite{Jimenez:2003} \\
    \hline
    0.17 & 83  $\pm$  8    & \cite{Simon:2005} \\
    0.27 & 77  $\pm$  14   & \\
    0.4  & 95 $\pm$ 17     & \\
    0.9  &117 $\pm$ 23     & \\
    1.3  &168 $\pm$ 17     & \\
    1.43 &177 $\pm$ 18     & \\
    1.53 &140 $\pm$ 14     & \\
    1.75 &202 $\pm$ 40     & \\
    \hline
    0.48 & 97  $\pm$  62   & \cite{Stern:2010} \\
    0.88 & 90  $\pm$  40   & \\
    \hline
    0.1791 &  75  $\pm$  4 & \cite{Moresco:2012}\\
    0.1993 & 75  $\pm$  5  & \\
    0.3519 & 83  $\pm$  14 & \\
    0.5929 & 104  $\pm$  13 & \\
    0.6797 & 92  $\pm$  8  & \\
    0.7812 & 105  $\pm$  12 & \\
    0.8754 & 125  $\pm$  17 & \\
    1.037  & 154  $\pm$  20 & \\
    \hline
    0.07   & 69  $\pm$  19.6   & \cite{Zhang:2014} \\
    0.12   & 68.6  $\pm$  26.2 & \\
    0.2    & 72.9  $\pm$  29.6 & \\
    0.28   & 88.8  $\pm$  36.6 & \\
    \hline
    1.363  & 160  $\pm$  33.6  & \cite{Moresco:2015}  \\
    1.965  & 186.5  $\pm$  50.4 & \\
    \hline
    0.3802 & 83  $\pm$  13.5   & \cite{Moresco:2016} \\
    0.4004 & 77  $\pm$  10.2   & \\
    0.4247 & 87.1  $\pm$  11.2 & \\
    0.4497 & 92.8  $\pm$  12.9 & \\
    0.4783 & 80.9  $\pm$  9    & \\
    \hline
    0.47   & 89  $\pm$  50  & \cite{Ratsimbazafy:2017} \\
    \hline
    1.26   & 135  $\pm$ 65  & \cite{Tomasetti:2023} \\
\hline
\end{tabular}
\end{table}

After improving their selection techniques and flux measurements, \cite{Risaliti:2019}
constructed a final sample of 1598 quasars with accurate measurements of the intrinsic 
UV and X-ray fluxes in the redsfhit range $0\lesssim z\lesssim 6$. These are the sources 
we shall use for the analysis in this paper, though only a portion of this redshift range 
will actually be matched to the more restricted redshift coverage (i.e., $z\lesssim 2$)
of the cosmic chronometers.

As noted above, we simultaneously optimize all of the quasar parameters, those
for $D_L$ in Equation~(\ref{eq:DLcosmo}) and $\tilde\beta$, $\gamma$ and $\delta$
in Equation~(\ref{eq:FXFUV}), using solely the quasar data on their own. This 
avoids any possible contamination from outside calibrators, and is in fact 
analogous to what one does with Type~Ia SNe, where the so-called `nuisance' 
parameters shaping the SN lightcurve are optimized along with parameters of 
the cosmological model itself. The results of this fitting are shown in Table~1, 
including the polynomial coefficients $a_2$ and $a_3$ in the cosmographic expression 
for $D_L$. 

Previous work has shown that $\gamma\sim 0.5-0.7$ when external sources are used to 
calibrate the quasar data
\citep{Avni:1986,Just:2007,Young:2010,Lusso:2010,Risaliti:2015,Lusso:2016,Risaliti:2019}.
A quick inspection of Table~1 shows that the optimization of $\gamma$ via the
use of internal calibration produces results fully consistent with this range. But as 
noted earlier, the slight differences between the values of these parameters optimized
this way compare to their estimation using external calibrators are sufficient 
to mitigate the tension---first pointed out by \cite{Risaliti:2019}---between the 
quasar Hubble diagram and the predictions of $\Lambda$CDM.

Note that the Hubble constant $H_0$ is not independent of $\tilde{\beta}$.
One can easily see this from Equation~(\ref{eq:FXFUV}), where $\tilde{\beta}$ and
$H_0$ combine to produce the `single variable' $\tilde{\beta}-2(\gamma-1)\log_{10}H_0$.
For the purpose of optimization, we have therefore subsumed $H_0$ into the parameter 
$\tilde{\beta}$. To make the results easy to interpret, however, we assume a fiducial 
value $H_0=70$ km s$^{-1}$ Mpc$^{-1}$, and the optimization for $\tilde{\beta}$ shown
in Table~1 corresponds to this choice. For a different Hubble parameter, $H_0^\prime$, 
the optimized value of $\tilde{\beta}$ in Table~1 would be changed by the amount
$\Delta{\tilde{\beta}}=2(\gamma-1)\log_{10}(H_0^\prime/H_0)$. Given the degeneracy
between $\tilde{\beta}$ and $H_0$, the actual value of the Hubble constant does not
affect the calibration of the UV-X-ray correlation function, i.e., it has no
impact on any of the other optimized variables, $\gamma$, $\delta$, $a_2$ and $a_3$. 
And to reiterate, $H_0$ completely cancels out from the righthand side of
Equation~(\ref{eq:dckappa}), so this analysis is independent of the contentious 
Hubble constant.

The expansion rate of the Universe, $H(z)$, and the integrated function
$\int_0^z du/H(u)$, are obtained directly from the redshift-time derivative 
$dz/dt$, using 
\begin{equation}
H(z)=-\frac{1}{1+z}\frac{dz}{dt}\;,\label{eq:Hzdiff}
\end{equation}
at any redshift $z\not=0$. The quantity $dz/dt$ may be measured from the differential 
age evolution of passively evolving galaxies, without the need to assume any particular
cosmological model \citep{Jimenez:2002,Jimenez:2003}. These galaxies are commonly
referred to as `cosmic chronometers.' A recent sample of 32 cosmic-chronometer 
measurements (see \citealt{Ruan:2019} and references therein) is shown in Table~2.

\begin{table}
\begin{center}
\caption{Speed of light in units of $c_0$ as a function of redshift}
\begin{tabular}{lll}
\hline\hline \\
Redshift & $c(z)/c_0$ & $\sigma_{c/c_0}$ \\
\\
\hline \\
0.07 & 0.991 & 0.002 \\
0.09 & 0.986 & 0.018 \\
0.12 & 0.977 & 0.060 \\
0.17 & 0.976 & 0.100 \\
0.18 & 0.980 & 0.100 \\
0.20 & 0.985 & 0.101 \\
0.20 & 0.985 & 0.101 \\
0.27 & 0.970 & 0.113 \\
0.28 & 0.971 & 0.114 \\
0.35 & 0.976 & 0.121 \\
0.38 & 0.963 & 0.122 \\
0.40 & 0.982 & 0.123 \\
0.42 & 0.967 & 0.123 \\
0.45 & 0.983 & 0.140 \\
0.47 & 0.970 & 0.125 \\
0.48 & 0.970 & 0.125 \\
0.59 & 0.971 & 0.136 \\
0.68 & 0.969 & 0.136 \\
0.78 & 0.963 & 0.138 \\
0.88 & 0.961 & 0.139 \\
0.90 & 0.943 & 0.121 \\
1.04 & 0.958 & 0.140 \\
1.26 & 0.958 & 0.140 \\
1.30 & 0.958 & 0.140 \\
1.36 & 0.958 & 0.140 \\
1.43 & 0.958 & 0.140 \\
1.53 & 0.955 & 0.141 \\
1.75 & 0.911 & 0.125 \\
1.97 & 0.910 & 0.126 \\
\\
\hline
\end{tabular}
\end{center}
\end{table}

\begin{figure}
\centering
\includegraphics[scale=0.48]{fig1.eps}
\caption{Distribution of $\kappa(z)\equiv c(z)/c_0$ values 
at $z=0.48$ based on 10,000 Monte Carlo simulations using the
measured uncertainties in the coefficients $\gamma$,
$\delta$, $\tilde\beta$, $a_2$, and $a_3$ and the
$1\sigma$ errors in Table~2. The dispersion corresponding
to the best fit Gaussian (solid black curve) is
$\sigma=0.133$.}
\label{fig1}
\end{figure}

\begin{figure}
\centering
\includegraphics[scale=0.48]{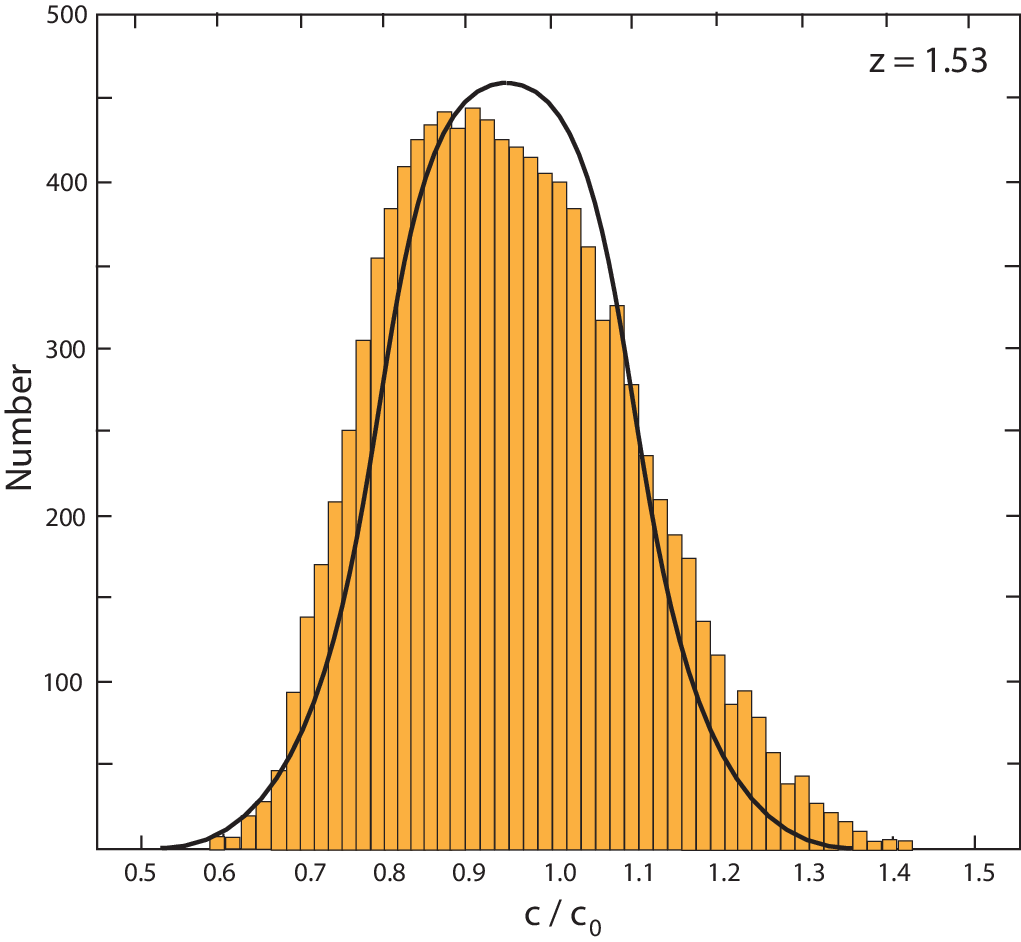}
\caption{Same as Fig.~1, except now at $z=1.53$.}
\label{fig2}
\end{figure}

\begin{figure}
\centering
\includegraphics[scale=0.53]{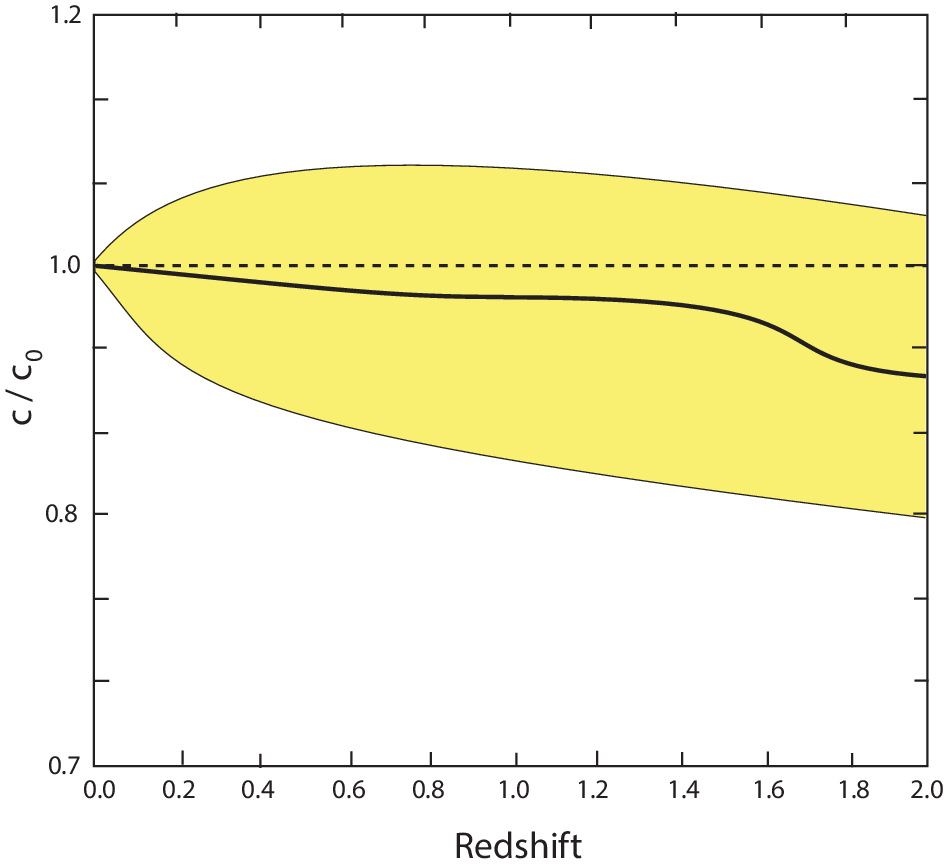}
\caption{{\sl Solid black curve:} the speed of light, $c(z)$, in units of 
$c$ (i.e., the normalized speed $\kappa(z)$ defined in the
text) as a function of redshift, calculated from the integration of 
Equation~(\ref{eq:dc}). {\sl Yellow shaded region:} the $\pm 1\sigma$ error 
estimated with 10,000 Monte Carlo simulations, as illustrated in Figs~1 
and 2. By comparison, the dashed line corresponds to $c(z)/c_0=1$. The 
measured value of $c(z)/c_0$ is fully consistent with zero variation
within the measurement errors.}
\label{fig3}
\end{figure}

We integrate Equation~(\ref{eq:dckappa}) over redshift, starting at
$z=0.09$ (the first entry in Table~2) where c(0) is assumed to 
have the value $c_0$, i.e., where $\kappa=1.0$, using the empirically derived quantities
$\tilde{D}_L$, $\tilde{D}_L^\prime$, $H(z)$ and $\int_0^z du/H(u)$. Estimating
the errors of the `measured' value of $\kappa$ at any given redshift
is complicated, however, in part because the errors incurred
with the integration of the function $1/H(z)$ over $z$ are 
correlated with the errors in $H(z)$ itself. To address this
difficulty, we instead estimate the error in $\kappa(z)$ at each
given value of $z$, based on 10,000 Monte Carlo simulations 
utilizing the measured uncertainties in the coefficients 
$\gamma$, $\delta$, $\tilde\beta$, $a_2$ and $a_3$ in Table~1, 
and the $1\sigma$ errors quoted in Table~2. All of these
variables are assumed to be distributed normally with dispersions
set equal to their reported errors.

In Figures~1 and 2 we show the resultant distributions in $\kappa(z)$
at two specific redshifts, $z=0.48$ and $1.53$. Though not perfectly
normal, these distributions are nevertheless matched quite well 
by Gaussian fits (solid curves), from which the $1\sigma$ errors
may be extracted. We find $\sigma_{c/c_0}\approx 0.125$ in both 
cases. Table~3 lists the value of $\kappa(z)$ calculated in this 
fashion throughout the redshift range $0\lesssim z\lesssim 2$.
As one may easily confirm, this ratio's deviation from $1$ is 
always well within the inferred $1\sigma$ error (see also Fig.~3). 
In other words, the speed of light estimated with this approach 
appears to be fully consistent with $c_0$ all the way out to 
redshift $\sim 2$, when the Universe was roughly one-third 
as old as it is today.

\section{Discussion}\label{discussion}
The use of quasars for cosmological analysis out to a redshift exceeding
$\sim 6$ has already been firmly established, notably for the optimization
of cosmological parameters in the standard model \citep{Risaliti:2019},
and for model selection between competing cosmologies, such as $\Lambda$CDM
and $R_{\rm h}=ct$ \citep{Melia:2019c}. The interesting suggestion to also
use them for the purpose of testing the constancy of $c$ over cosmological
distances \citep{Liu:2023} has pointed to a possible anomaly, however, 
revealing a $\sim 2.7\sigma$ deviation of the speed of light from its 
value measured on Earth.

This tension mirrors a similar inconsistency identified by \cite{Risaliti:2019}
between the inferred quasar Hubble diagram and the predictions of $\Lambda$CDM.
What is more puzzling, though, is that the variation of $c(z)/c_0$ identified
by \cite{Liu:2023} is based on the use of a cosmographic polynomial fit, not
directly related to $\Lambda$CDM. In principle, this tension would thus appear
to be more general, not tied to any particular choice of background cosmology.

But though our previous comparative test between $\Lambda$CDM and $R_{\rm h}=ct$
concluded that the quasar data favour the latter over the former, we also
demonstrated that the tension between these data and the standard model is largely
mitigated when one uses internal calibration (via the simultaneous optimization
of all the parameters) instead of external sources, such as Type~Ia supernovae.
It also appears that an incorrect relation between $D_L$ and $H(z)$ was used
in the previous measurement of $c(z)/c_0$, given that terms involving $dc/dz$
cannot be ignored when the outcome points to a variable $c(z)$.

We have thus retested the presumed constancy of $c$ over cosmological distances
using the high-$z$ quasar observations together with the expansion rate $H(z)$
inferred from cosmic chronometers, using the alternative calibration of the
quasar data and an updated expression for $d(c/c_0)/dz$. We now find no evidence
for a variation of $c$, based on these observations, all the way out to $z\sim 2$.

It is important to note that this outcome confirms the result already
discussed by \cite{Liu:2023} based on the use of Type Ia supernova data. In both
cases, the profile of $c(z)$ with redshift suggests no variation over cosmic
scales and times. This conclusion is relevant to the larger question of how
confident we should be in the use of the FLRW metric to describe the cosmic
spacetime as opposed to possible alternative interpretations of cosmic redshift 
based on the assumption that $c$ has varied in our past 
\citep{Albrecht:1999,Barrow:1999a,Bassett:2000,Moffat:2002}. 

Thus, in addition
to these conclusions based on the use of Type Ia SNe and quasars, it would be
very useful to redo the analysis utilizing Equations~(\ref{eq:FLRW})--(\ref{eq:dc})
with other classes of sources, particularly yielding other measures of distance
and/or age. For example, strong gravitational lenses provide us with the means
of `measuring' the ratio of angular-diameter distances to the lens and background
source (see, e.g., \citealt{Wei:2020b}). A clear benefit of this approach is the
elimination of $H_0$ from the analysis, which avoids the current uncertainty with
its value. When combined with the cosmic chronometer measurements of $H(z)$, these
sources should provide an important affirmation of the constancy of $c$ over
cosmic distances analogously to the work reported in \cite{Liu:2023} and in this
paper. A similar analysis may also be feasible with the use of HII galaxies to
construct the Hubble diagram \citep{Yennapureddy:2017}, from which the redshift
dependence of $c$ may be inferred along the lines described in this paper.
This work is underway and its results will be reported elsewhere. 

\section{Conclusion}\label{conclusion}
In closing, we reiterate the very important feature of this work that it is
completely independent of the value of $H_0$. This is crucial because it now 
appears that we do not have a clear understanding of how this parameter is 
to be measured most accurately \citep{Riess:2021}. In addition, our use of 
a cosmographic fit for $D_L$ and $D_L^\prime$ has rendered these results 
completely free of any presumed cosmological model, other than their generic 
dependence on the viability of the FLRW metric. It thus appears that the 
assumption of a constant speed of light, required for the derivation of the
FLRW spacetime in the first place, is fully consistent with all of the data we 
have at our disposal today.

\section*{Acknowledgments}
I am very grateful to Beta Lusso and Guido Risaliti for sharing their
data for this analysis. I am also grateful to the anonymous
referee for useful comments.

\section*{DATA AVAILABILITY STATEMENT}
No new data were generated or analysed in support of this research.

\bibliographystyle{mn2e.bst}
\bibliography{ms.bib}

\label{lastpage}
\end{document}